\documentclass[12pt,twoside]{article}
\usepackage[english]{babel}
\usepackage{graphicx,rotating,epsfig}
\usepackage{a4p}
\usepackage{xspace}

\begin{document}


\newcommand{\fix}[1]{} 
\newcommand{\MET}{\mbox{$\raisebox{.3ex}{$\not\!$}E_T$}}
\newcommand{\METVEC}{\mbox{$\raisebox{.3ex}{$\not\!$}{\vec E}_T$}}
\newcommand{\MPTVEC}{\mbox{$\raisebox{.3ex}{$\not\!$}{\vec {p}_T}$}}
\newcommand{\ttbar}     {\mbox{$t\bar{t}$}}
\def\DZero{D0~}

\newcommand{\Et}{\mbox{$E_T$}}
\newcommand{\Pt}{\mbox{$p_T$}}
\newcommand{\Ht}{\mbox{$H_T$}}
\newcommand{\met}{\mbox{$\protect \raisebox{0.3ex}{$\not$}\Et$}}
\newcommand{\mpt}{\mbox{$\protect \raisebox{0.3ex}{$\not$}\Pt$}}

\begin{center}
  {\LARGE FERMI NATIONAL ACCELERATOR LABORATORY}
\end{center}

\begin{flushright}
       FERMILAB-TM-2440-E \\
       CDF Note 9870 \\
       D\O\ Note 5973 \\
       August 2009 \\
\end{flushright}

\vskip 1cm

\begin{center}

 {\LARGE\bf
    Combination of CDF and D0 Measurements \\ \vspace{0.1in}
	of the Single Top Production Cross Section
 }

  \vfill
 {\Large
    The Tevatron Electroweak Working Group \footnote{The Tevatron Electroweak
    Working Group can be contacted at tev-ewwg@fnal.gov.\\
    \hspace*{0.20in} More information can
    be found at {\tt http://tevewwg.fnal.gov}.} \\
    for the CDF and D\O\ Collaborations\\
  }
\end{center}


\begin{abstract}
We report a combination of the CDF and D0 measurements of the inclusive single top quark production
cross section in the $s$- and $t$-channels, $\sigma_{s+t}$, in $p{\bar{p}}$ collisions
at a center of mass energy of 1.96~TeV.  The total integrated luminosity included in CDF's 
analysis is 3.2~fb$^{-1}$ and D0's analysis
has 2.3~fb$^{-1}$.  A Bayesian analysis is used to extract the cross section from the
distributions of multivariate discriminants provided by the collaborations.
For a top quark mass $m_t=170$~GeV$\!/c^2$, we measure a
cross section of $2.76^{+0.58}_{-0.47}$~pb. We extract
the CKM matrix element $|V_{tb}|=0.88 \pm 0.07$ 
with a 95\% C.L. lower limit of $|V_{tb}|>0.77$.  
\end{abstract}

\vfill







\section{Introduction}


In the standard model (SM), top quarks are expected to be produced 
singly in $p{\bar p}$ collisions through
$s$-channel or $t$-channel exchange of a virtual $W$ boson~\cite{willen}.
The reasons for studying single top quarks are
compelling: the production cross section is directly proportional to
the square of the CKM matrix~\cite{ckm} element $|V_{tb}|$, and the
cross section measurement is sensitive to fourth-generation models, models
with flavor-changing neutral currents, and other new
phenomena~\cite{Tait:2000sh}. The main backgrounds to the single top
final state are $W$+jets production, \ttbar~ production, and QCD multijet production.
Electroweak production of single top
quarks is a difficult process to measure because the expected
production cross section ($\sigma^{\rm theory}_{s+t}\sim2.9-3.5$~pb~\cite{harris,stxs})
is much smaller than those of the background processes, and it is
also smaller than the uncertainty on the total background.
The presence of only one top quark in the event provides fewer
features to use in separating the signal from background, compared
with measurements of top pair production (\ttbar), which was
first observed in 1995~\cite{ttbar}.  


The CDF and D0 collaborations have developed a variety of sophisticated multivariate
techniques that take advantage of the differences in the kinematic distributions between
signal and backgrounds.  These have successfully been used to observe the process, at the 
5.0 standard deviation level of significance, separately by D0~\cite{d0obsprl} and by
CDF~\cite{cdfobsprl}.  Further details of these analyses
are given in earlier publications presenting evidence for single top production
in smaller samples of data~\cite{d0evidenceprl,d0evidenceprd,cdfevidenceprl}.

Because the presence of single top production has been firmly established by the two
collaborations and due to the computational difficulty of computing the significance
of the two in combination, we do not perform a hypothesis test of the combined results.
The cross section measurements, however, remain limited in precision by the sizes of the
data samples collected by the two collaborations, and thus a combination of the results
provides a more precise measurement and also stronger constraints on the CKM
matrix element $|V_{tb}|$.


\section{Method}

The CDF and D0 collaborations both measure the single top cross section and extract
$|V_{tb}|$ using a Bayesian statistical analysis. We use the same method here to combine the 
two measurements. 
The CDF and D0 collaborations use independently developed computer
programs for performing the calculation,
and we have verified that both codes reproduce the individual experiment results. 
Since the two codes produce consistent results we use the D0 code here to obtain the 
combined Tevatron results.
CDF and D0 have exchanged histograms of their final multivariate discriminants 
for the observed data and all signals
and backgrounds in each analysis channel. Furthermore, all systematic uncertainties 
-- those on the predicted rates, histogram shapes, and bin-by-bin independent Monte Carlo 
statistical uncertainties, have been exchanged as well.

Rate and shape uncertainties are categorized by their
source for use in evaluating correlations, between signals and/or backgrounds within a 
channel, between channels, and between experiments.
Sources of systematic uncertainty that are common to the measurements of the two collaborations
have been assigned 100\% correlation, and other sources of uncertainty are assumed to be
uncorrelated.  A list of systematic uncertainties is given
in Table~\ref{tab:tevsyst} and discussed in more detail below. 

The method used for extraction of the cross section chosen by the collaborations is a Bayesian
technique~\cite{bayes-limits}. The posterior probability density is given by
\begin{equation}
p(\sigma_{s+t}) = \int L({\rm data}|\sigma_{s+t},\{\theta\})\pi(\sigma_{s+t})
\pi(\{ \theta \}) d\{\theta\} \, , \label{eq:posterior}
\end{equation}
where $L$ is the joint likelihood function of all bins in all histograms
\begin{equation}
L = \prod_{bins,\,channels} \frac{(s_i + b_i)^{n_i} e^{-(s_i+b_i)}}{n_i!},
\end{equation}
$\{\theta\}$ is the set of nuisance parameters (such as integrated luminosity, $b$-tag efficiency, 
and other rate and shape parameters),
and $\pi(\{ \theta \})$ is the product of the priors encoding the systematic uncertainties 
on $\{ \theta \}$.  
The predictions for the number of signal events ($s_i$) and background events ($b_i$) 
depend on the values of the nuisance parameters.

The prior density for the signal cross section is taken to be a nonnegative
flat prior, $\pi(\sigma_{s+t}) = 1/\sigma_{\rm max}$ for $\sigma \ge 0$, and
$\pi(\sigma) = 0$ otherwise. Here, $\sigma_{\rm max}$ is the highest signal cross section
axis point (7~pb for the combination).
The priors on $\{\theta\}$ are truncated Gaussians, where the truncations are introduced 
so as not to allow negative predictions of $s_i$ or $b_i$ for any template in any bin. 
The integral over $d\{\theta\}$ is performed through numerical integration.

We quote the measured cross section as the value that maximizes the
posterior likelihood, and the uncertainty as the shortest interval containing 68\% of
the posterior area. 

In addition to the above, the D0 code uses a Gamma prior for the MC statistics uncertainty, 
reflecting the Poisson nature of MC statistics, and having the additional advantage
that an analytic solution to the integral in Eq.~\ref{eq:posterior} exists. This does not
bias the cross section but reduces the computing time considerably.

The combination procedure includes all channels from both collaborations in 
the Bayesian analysis, including correlations of systematic uncertainties between
experiments. We combine the CDF super-discriminant, 
the CDF missing transverse energy (\met) plus jets analysis, and the D0 combination
discriminant.
While there are some differences in the detailed treatments of uncertainties, 
the CDF and D0 codes give results consistent with each other when 
run over the individual inputs. We have verified this by running the D0 code over
the CDF inputs.

\begin{table}[!h!btp]
\caption{Results summary for CDF's $\ell$+jets super-discriminant analysis, 
\met+jets analysis, and CDF's total combination, comparing the CDF code to the D0 
code. This is for the published CDF measurements at $m_t=175$~GeV$\!/c^2$.
}
\begin{center}
\begin{tabular}{lcc} \hline
Analysis             & Cross Section (pb)  & Cross Section (pb) \\
                     & CDF Calculation     & D0 Recalculation \\ \hline
\multicolumn{3}{l}{No systematic uncertainties} \\
~~Super-discriminant & 2.29$^{+0.46}_{-0.43}$  & 2.29$^{+0.45}_{-0.43}$       \\
~~\met+jets          & 4.18$^{+1.43}_{-1.35}$  & 4.19$^{+1.41}_{-1.37}$      \\
~~Combined           & 2.49$^{+0.43}_{-0.43}$  & 2.48$^{+0.44}_{-0.42}$         \\ 
\multicolumn{3}{l}{Including systematic uncertainties} \\
~~Super-discriminant & 2.1$^{+0.6}_{-0.5}$ & 2.1$^{+0.6}_{-0.5}$       \\
~~\met+jets          & 4.9$^{+2.5}_{-2.2}$ & 5.2$^{+2.3}_{-2.1}$      \\
~~Combined           & 2.3$^{+0.6}_{-0.5}$ & 2.3$^{+0.6}_{-0.5}$         
\end{tabular}
\end{center}
\label{tab:cdfresults}
\end{table}

\begin{table}[!h!btp]
\caption{Results summary for CDF's $\ell$+jets super-discriminant analysis, 
\met+jets analysis, and CDF's total combination, comparing the CDF code to the D0 
code. This is for the re-calculation corresponding to $m_t=170$~GeV$\!/c^2$ and
includes all systematic uncertainties.
}
\begin{center}
\begin{tabular}{lcc} \hline
Analysis        & Cross Section (pb)  & Cross Section (pb) \\
                & CDF Calculation     & D0 Recalculation \\ \hline
~~Super-discriminant & 2.17$^{+0.56}_{-0.55}$   & 2.12$^{+0.59}_{-0.54}$       \\
~~\met+jets          & 5.0$^{+2.6}_{-2.3}$      & 5.4$^{+2.8}_{-2.7}$      \\
~~Combined           & 2.35$^{+0.56}_{-0.50}$   & 2.29$^{+0.60}_{-0.54}$         
\end{tabular}
\end{center}
\label{tab:cdfresults170}
\end{table}

 
Table~\ref{tab:cdfresults} (Table~\ref{tab:cdfresults170}) lists the measured cross sections 
with and without systematic uncertainties for the two ingredients to the CDF combination as 
well as the CDF combination result, both for the original CDF calculation and the recalculation 
using the D0 code, for a top quark mass of $m_t=175$~GeV$\!/c^2$ ($m_t=170$~GeV$\!/c^2$). 
The D0 code reproduces
the CDF results, indicating that all inputs and uncertainties are being read properly. 
%

Different conventions have been used by the CDF and D0 analyses for the top quark mass
and the theoretical single top cross section. CDF assumes $m_t=175$~GeV$\!/c^2$ 
in its extraction of the
central value of the single top production cross section, while D0 assumes $m_t=170$~GeV$\!/c^2$. 
The current world average top mass is $173.1 \pm 1.3$~GeV$\!/c^2$~\cite{tevtopmass}.
CDF includes in its $p$-value calculation and the $|V_{tb}|$ limit the uncertainty in $m_t$ 
and thus all rate and shape uncertainties have been evaluated for the single top signal 
templates and the $t{\bar t}$ background templates.
D0 does not have single top or \ttbar~ samples for $m_t=175$~GeV$\!/c^2$, 
so CDF has shifted its predictions to $m_t=170$~GeV$\!/c^2$, and thus the results quoted here 
assume $m_t=170$~GeV$\!/c^2$. The discriminant shapes and normalizations
have been shifted for s-channel and t-channel signal as well as for \ttbar~background. The
\ttbar~background cross section at $m_t=170$~GeV$\!/c^2$ is $7.91 \pm 0.95$~pb~\cite{ttbar-xsec}.
Other background normalizations that weakly depend on \ttbar~ have not been modified. 
This has a small impact on the measured cross section as can be seen in 
Table~\ref{tab:cdfresults170}.

The other difference in convention is the choice of theoretical cross section to compare the 
measurement with. CDF chooses Harris and Sullivan's next-to-leading order (NLO) calculation 
with cross sections of
$\sigma^{\rm theory}_s=0.99 \pm 0.07$~pb and $\sigma^{\rm theory}_t=2.15 \pm 0.24$~pb at 
$m_t=170$~GeV$\!/c^2$~\cite{harris}, while D0 chooses Kidonakis' NLO plus soft-gluon 
corrections calculation with cross sections of $\sigma^{\rm theory}_s=1.12 \pm 0.05$~pb and 
$\sigma^{\rm theory}_t=2.34 \pm 0.13$~pb~\cite{stxs}. This choice has no effect on the 
measured cross section, but it does affect the extraction of $|V_{tb}|$ and its limits.  
We report both extractions of $|V_{tb}|$ here. 


\section{Systematic Uncertainties}

The complete set of systematic uncertainties and their correlations from the CDF and D0 
analyses have been included in the combination. They are categorized in Table~\ref{tab:tevsyst}. 

The systematic error categories in Table~\ref{tab:tevsyst} mainly follow those used for the 
Tevatron top mass combination~\cite{tevtopmass} aiming to lump together sources of systematic 
uncertainty that share the same or similar origin as well as the same correlation between the 
two experiments. Each error category is briefly described below.

\begin{description}

\item[Luminosity from detector:] That part of the luminosity uncertainty that comes from the 
uncertainty on the luminosity detector acceptance and efficiency. It is uncorrelated between 
CDF and D0.

\item[Luminosity from cross section:] That part of the luminosity uncertainty that comes from 
the uncertainty of the inelastic and diffractive cross sections. It is correlated between 
CDF and D0.

\item[Signal modeling:] The systematic uncertainty arising from uncertainties in the modeling 
of the single top signal. This includes uncertainties from variations in the ISR, FSR, and PDF 
descriptions. It also includes difference in the hadronization models. It is correlated between 
CDF and D0.

\item[Background from MC:] The systematic uncertainty arising from uncertainties in modeling 
of the different background sources that are correlated between CDF and D0. It includes \ttbar\ 
and diboson normalization uncertainties obtained from theoretical calculations. 

\item[Background from data:] The systematic uncertainty arising from uncertainties in modeling 
of the different background sources that are obtained using data-driven methods and are
uncorrelated between CDF and D0. It includes the 
uncertainty on the $W$+jets and $Wbb/Wcc$ normalization, scale factor and shape  
as well as the uncertainty on the multijet modeling and normalization.
It also includes the uncertainty due to MC statistics.

\item[Detector modeling:] The systematic uncertainty arising from the uncertainty 
on the event detection efficiencies for object identification and 
MC mismodeling of data. It is uncorrelated between CDF and D0.

\item[b-tagging:] The systematic uncertainty coming from the uncertainty on the b-tagging and 
mistag rate and shape modeling. It is uncorrelated between CDF and D0.



\item[dJES:] The part of the JES uncertainty which originates from
   limitations in the calibration data samples used.  
   For CDF this corresponds to uncertainties associated
   with the $\eta$-dependent JES corrections which are estimated
   using di-jet data events. For D0 this includes uncertainties in
   the calorimeter response for light jets, uncertainties from
   $\eta$- and $p_{T}$-dependent JES corrections, and other small contributions.
   It is uncorrelated between CDF and D0.

\end{description}

\begin{table}[!h!btp]
\caption{Sources of systematic uncertainty in the CDF and D0 single top analyses,
together with the rate range for each uncertainty and whether the shape effect on
the final discriminant is considered for each uncertainty or not. }
\begin{center}
\begin{tabular}{l|cc|cc|c}\hline
Systematic Uncertainty          & \multicolumn{2}{c|} {CDF} &  \multicolumn{2}{c|} {D0} & Correlated between \hspace{-0.2in}\\ 
                                &               &               &               &       & the two experiments \hspace{-0.2in} \\ \hline
                                & Rate          & \hspace{-0.1in}Shape         & Rate          & \hspace{-0.1in}Shape & \\ \hline 
Luminosity from detector        & 4.5\%         &               & 4.6\%         &       & \\ 
Luminosity from cross section\hspace{-0.05in}   & 4.0\%         &               & 4.0\%         &       & $\bullet$ \\ 
Signal modeling                 & 2.2--19.5\%   & $\bullet$     & 3.5--13.6\%   &       & $\bullet$ \\
Background from MC              & 12.1--12.4\%  & $\bullet$     & 15.1 \%       &       & $\bullet$ \\
Background from data            & 17--40\%       & $\bullet$     & 13.7--54\%  & $\bullet$     & \\
Detector modeling               & 0--9\%        & $\bullet$     & 7.1 \%        &       & \\
b-tagging                       & 0--29\%      & $\bullet$     & 2--30\%   & $\bullet$ & \\ 
dJES                            & 0-16\%        & $\bullet$     & 0.1--13.1\%   & $\bullet$ & \\

\end{tabular}
\end{center}
\label{tab:tevsyst}
\end{table}

We do not include the small uncertainty due to the top quark mass as our result is computed at
$m_t=170$~GeV$\!/c^2$. We also do not include the systematic uncertainty on the single top cross 
section theory prediction when measuring the cross section, but we do include it as an 
additional systematic when extracting $|V_{tb}|$.

\clearpage

\section{Results}

\begin{figure}[!h!tbp]
\begin{center}
\includegraphics[width=0.48\textwidth]{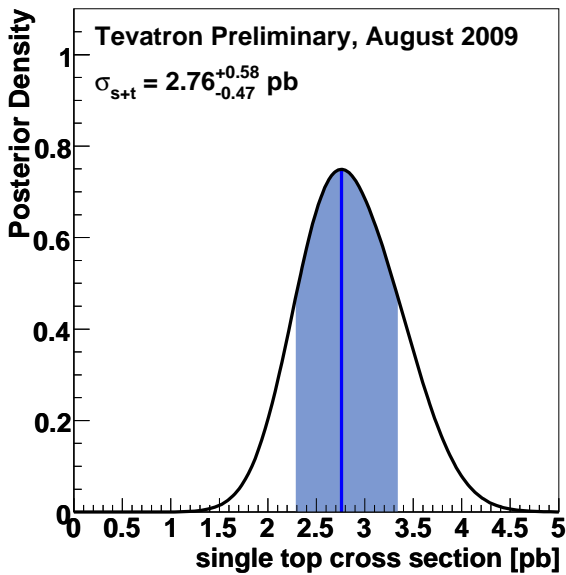}
\caption{The posterior probability distribution of the combined CDF and D0 analyses.}
\label{fig:tevposterior}
\end{center}
\end{figure}

Figure~\ref{fig:tevposterior} shows the posterior probability distribution of 
the combined CDF and D0 analyses, and Table~\ref{tab:summary} lists the measured
cross section.

\begin{table}[ht!]
\caption{Measured single top cross section and $|V_{tb}|$ value and lower limit
for two different theory cross sections. }
\begin{center}
\begin{tabular}{lc} 
\hline \hline \vspace{-0.15in} \\ 
measured cross section                 & $2.76^{+0.58}_{-0.47}$~pb \\
$|V_{tb}|$, $\sigma^{\rm theory}_{s+t}=3.46$~pb     & $0.88 \pm 0.07$  \\
$|V_{tb}|$ 95\% C.L. limit, $\sigma^{\rm theory}_{s+t}=3.46$~pb & 0.77 \\
$|V_{tb}|$, $\sigma^{\rm theory}_{s+t}=3.14$~pb     & $0.91 \pm 0.08$  \\
$|V_{tb}|$ 95\% C.L. limit, $\sigma^{\rm theory}_{s+t}=3.14$~pb & 0.79 \\
\hline \hline
\end{tabular}

\end{center}
\label{tab:summary}
\end{table}

The most probable value of the combined $s$-channel and $t$-channel cross
sections is $2.76^{+0.58}_{-0.47}$~pb for a top quark mass of 170~GeV$\!/c^2$. This value 
lies in between the CDF measurement of 2.35$^{+0.56}_{-0.50}$~pb and the D0 measurement
of $3.94 \pm 0.88$~pb as expected. The combination improves the cross section uncertainty
from about 22\% for each experiment to 19\% for the combination. We have also computed the 
compatibility of the two measurements with each other, based on
the uncorrelated systematic uncertainties, and find that they are 1.6 standard deviations
apart.

In order to extract information about $|V_{tb}|$ from
the combined data, we assume that the off-diagonal CKM matrix elements
$|V_{ts}|$ and $|V_{td}|$ are much smaller than $|V_{tb}|$, but we do not make assumptions about
the unitarity of the $3\times 3$ CKM matrix, allowing for a fourth generation of quarks.
We obtain the $|V_{tb}|^2$ posterior by dividing the measured cross section by the
theoretical single top cross section for $V_{tb}=1$ and a top quark mass of 170~GeV$\!/c^2$.

\begin{figure}[!h!tbp]
\begin{center}
\includegraphics[width=0.48\textwidth]{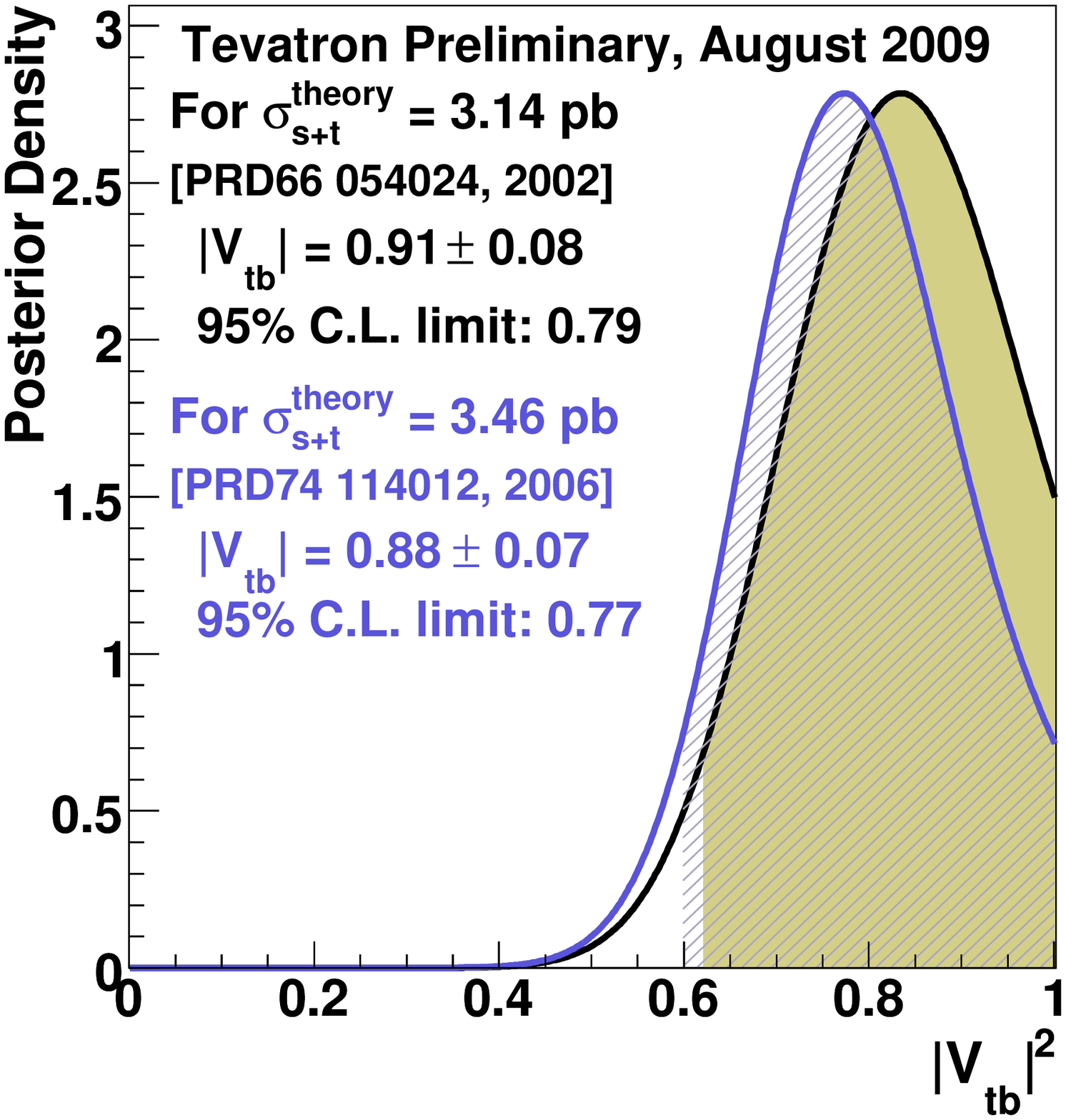}
\includegraphics[width=0.48\textwidth]{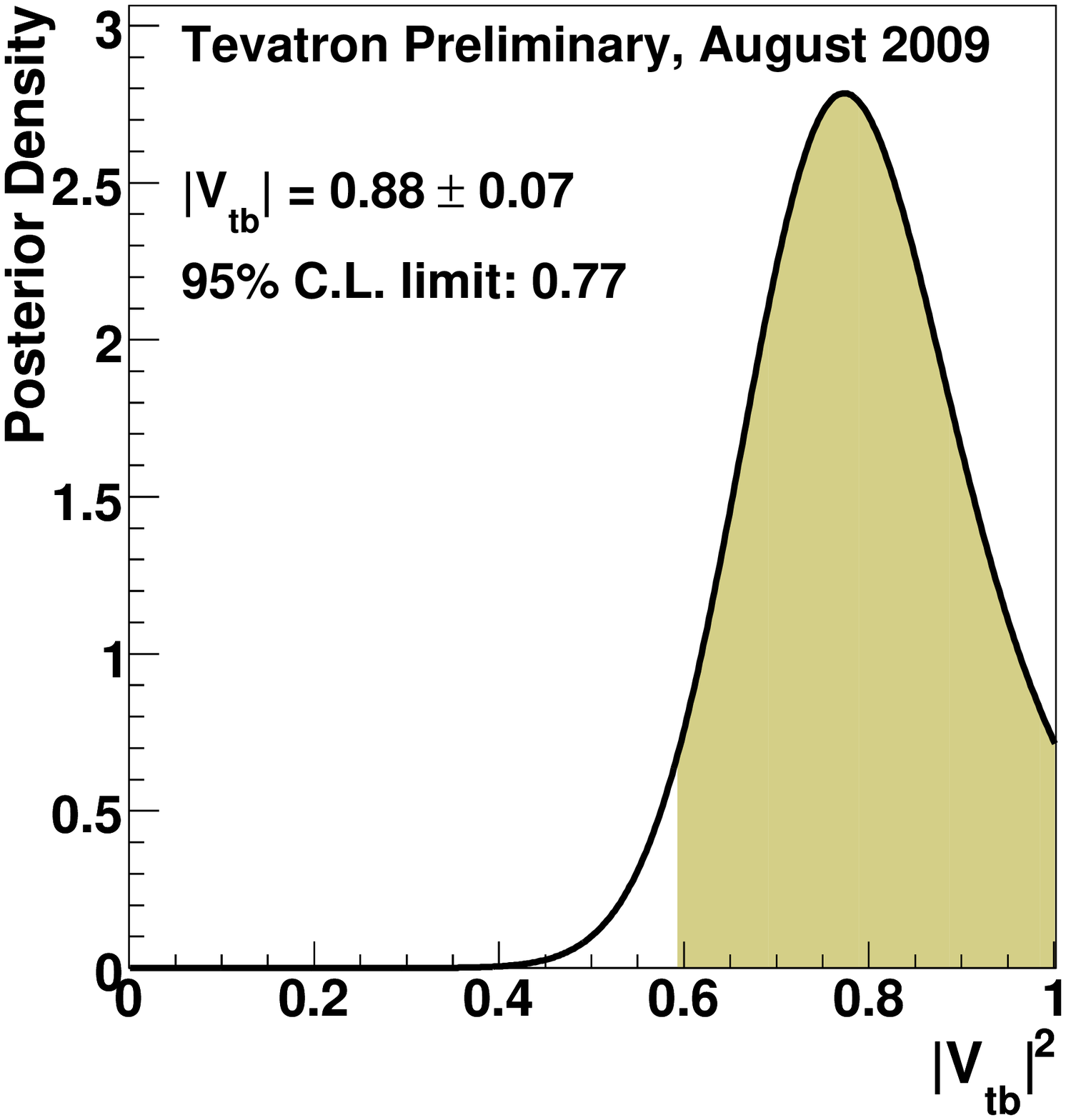}
\caption{The posterior probability distribution of the combined CDF and D0 analyses
for $|V_{tb}|$. The right plot shows only the posterior, $V_{tb}$ measurement and limit
using a theory cross section of 3.46~pb.}
\label{fig:tevvtbposterior}
\end{center}
\end{figure}

We obtain $|V_{tb}|=0.91 \pm 0.08$ using the theoretical calculation
of $\sigma^{\rm theory}_{s+t}=3.14 \times |V_{tb}|^2$~pb of Harris and Sullivan, 
and $|V_{tb}|=0.88 \pm 0.07$ using the theoretical
calculation $\sigma^{\rm theory}_{s+t}=3.46 \times |V_{tb}|^2$~pb of Kidonakis. We limit
$|V_{tb}|>0.79$ at the 95\% C.L. assuming a flat prior in $|V_{tb}|^2$
from 0 to 1 using the calculation of Harris and Sullivan, and limit
 $|V_{tb}|>0.77$ at the 95\% C.L. using the calculation of Kidonakis. The two
$|V_{tb}|$ posteriors are shown in Figure~\ref{fig:tevvtbposterior}. 
In comparison, the CDF extraction of $|V_{tb}|$ at 175~GeV$\!/c^2$ is $0.91 \pm 0.14$ and
the D0 extraction of  $|V_{tb}|$ at 170~GeV$\!/c^2$ is $1.07 \pm 0.12$. The relative
$|V_{tb}|$ uncertainty improves from 14\% (CDF) and 11\% (D0) to 8\% for the
combination, more than for the cross section due to the square root dependence.


\section{Summary}

\begin{figure}[!h!tbp]
\begin{center}
\includegraphics[width=0.6\textwidth]{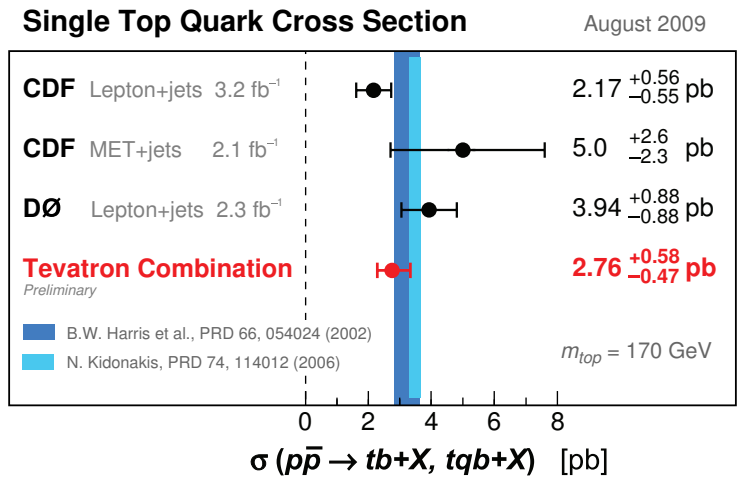}
\caption{Tevatron single top cross section measurements and their combination.}
\label{fig:summary}
\end{center}
\end{figure}

We combine the CDF and D0 measurements of the single top production
cross section $\sigma_{s+t}$ including the effects of all correlated and
uncorrelated systematic uncertainties.  The combined
measurement is $\sigma_{s+t} = 2.76^{+0.58}_{-0.47}$~pb, in agreement with the SM
expectation. A summary of the cross section measurements is shown in
Figure~\ref{fig:summary}. From this we obtain measurements
of $|V_{tb}|$ using two different theory cross section normalizations for a top 
quark mass of 170~GeV$\!/c^2$. We find $|V_{tb}|=0.88 \pm 0.07$ for a cross section
of $\sigma^{\rm theory}_{s+t}=3.46 \times |V_{tb}|^2$~pb and set a 95\% C.L. lower limit
of $|V_{tb}|>0.77$.

\end{document}